\documentclass[journal=jacsat,manuscript=article]{achemso}
\usepackage{orcidlink}
\usepackage[version=3]{mhchem} 
\usepackage{upgreek}
\usepackage{color}

\usepackage[normalem]{ulem}

\author{Pu Zhang\,\orcidlink{0000-0002-6253-0555}}
\affiliation{School of physics, Huazhong University of Science and Technology, Luoyu Road 1037, Wuhan 430074, China}
\email{puzhang0702@hust.edu.cn}

\author{Christos Tserkezis\,\orcidlink{0000-0002-2075-9036}}
\affiliation{POLIMA---Center for Polariton-driven Light--Matter Interactions, University of Southern Denmark, Campusvej 55, DK-5230 Odense M, Denmark}
\email{ct@mci.sdu.dk}

\author{N. Asger Mortensen\,\orcidlink{0000-0001-7936-6264}}
\affiliation{POLIMA---Center for Polariton-driven Light--Matter Interactions, University of Southern Denmark, Campusvej 55, DK-5230 Odense M, Denmark}
\altaffiliation{Danish Institute for Advanced Study, University of Southern Denmark, Campusvej 55, DK-5230 Odense M, Denmark}
\email{asger@mailaps.org}

\title[]
  {Quantum-hydrodynamic modal perspective on plasmonic gap structures}

\begin{document}

\begin{abstract}
Plasmonic gap structures are among the few configurations capable of generating extreme light confinement,
finding applications in surface-enhanced
spectroscopy, ultrasensitive detection, photocatalysis and
more. Their plasmonic response undergoes a dramatic, quantum
effect-driven transition as the gap size approaches zero. Modal
analysis can reveal insights into the mechanisms governing
this process, which are otherwise obscured by nonlocal damping
effects. Here, we offer a fresh modal perspective on the transition
of the plasmonic response using quantum hydrodynamic theory
(QHT)-based quasinormal mode (QNM) analysis. Focusing on the bonding
dipolar and charge-transfer plasmons of a nanosphere dimer, we
examine the detailed mode transition through the touching regime
as well as the asymptotic behavior compared with the classical results
as the constituent nanoparticles either separate or overlap.
The complex eigenfrequency particularly provides accurate
information on the linewidth and quality factor of the plasmon
modes. We introduce an index to characterize charge-transfer
efficiency, especially for the charge-transfer plasmon.
The significant role of nonlocal damping in the mode evolution is
elucidated by our mode-resolved QHT-QNM analysis. The insights from
our theoretical study provide an integrated understanding of mode
evolution in plasmonic gap structures, which can further
advance gap structure-based applications.
\end{abstract}

\section{Introduction}
Surface plasmons are collective oscillations of free electrons in
a conducting material driven by an optical field. The surface wave
behavior tightly localizes the optical field at the interface, bringing
about profound physical effects in light confinement and optical processes~\cite{Brongersma2010Plasmonics,Halas:2011,Baumberg2019ExtremeNanophotonics}.
The nanoscale confinement breaks the diffraction limit~\cite{Gramotnev:2010},
while the enhanced optical field significantly boosts a myriad of
light--matter interactions~\cite{xu_pre62,Halas:2011,acuna_sci338,Yu:2019}.
The modification of the local density of electromagnetic states by
surface plasmons can alter photon emission processes~\cite{Goncalves:2020,Chen2017},
while energetic hot electrons generated through these interactions actively
participate in chemical
reactions~\cite{brongersma_natnano10,li_nanoph6,khurgin_faraday214}.
The rich physics enabled by light confinement forms the cornerstone of
plasmonics research. In pursuit of stronger confinement, nanostructured
metals are highly effective. Metallic nanoparticles (NPs) provide
three-dimensional (3D) confinement through resonant localized surface
plasmons (LSPs)~\cite{le_nn2}. Sharp tips may concentrate the optical field through
a non-resonant lightning-rod effect~\cite{Stockman2004Nanofocusing}.
Plasmonic gap structures achieve the highest degree of confinement due to
both geometric compression and resonant effects~\cite{Hecht2012AtomicScale}.
Leveraging plasmonic gap structures enables the study of the aforementioned intriguing
physical effects, spawning far-reaching applications such as subwavelength
optical microscopy~\cite{Gramotnev:2010}, single-molecule detection through
surface-enhanced Raman spectroscopy~\cite{Ding:2016}, bright photon sources
due to plasmonic enhancement~\cite{Fernandez-Dominguez:2018}, and plasmonic
photocatalysis~\cite{Yuan:2023}, to name a few.

Rapid advances in nanofabrication and nanoscale manipulation techniques
have enabled the experimental realization of plasmonic gap structures
with increasingly smaller gap sizes and finely controlled
morphologies~\cite{Wang2013,Lee2016,Manfrinato2017,Kim2020}.
This aligns with the pursuit of achieving tighter light confinement and enhanced
device performance through increased plasmonic effects. Single-digit-nanometer
and subnanometer plasmonic gap structures not only provide further qualitative enhancements but also raise important questions about the ultimate limits of 
plasmonic structures~\cite{romero_oex14,Crozier2016,esteban_natcom3,liu_nn13,raza_ol15}.
The latter has sparked intense explorations of optical processes with
record-breaking plasmonic enhancement factors. For instance, subnanometer
resolution has been demonstrated with tip-enhanced spectroscopy~\cite{Dong2020photoluminescence,Apkarian2019vibrationalmodes}, where
the mode volume of the localized optical field enters the sub-1\,nm$^3$
regime~\cite{benz_sci354,Li2021,Wu2021}, and an average
field enhancement of around 10$^3$ is corroborated~\cite{Xu2024}. As the gap size approaches
the characteristic length scale of electron screening, quantum-related
phenomena emerge, potentially expanding or in some cases limiting the
application scenarios of plasmonics~\cite{Mortensen:2021,stamatopoulou_omex12}.
Inelastic electron tunneling enables electrically-driven plasmon
sources~\cite{Hecht2015,Greffet2016}, strong electron--photon
coupling facilitates ultrafast electro-optic
modulation~\cite{Li2021BiasModulation,Zurak2024}, and quantum tunneling
can trigger nonlinear optical processes~\cite{Borisov2012}. In this regard, the exploitation
of nearly-touching gap structures for innovations requires an
in-depth understanding of how the plasmonic response evolves as the gap size approaches zero.

To investigate the transition of the response of plasmonic gap
structures as the gap size decreases, a theoretical approach is
indispensable. Mode hybridization is a conceptually illuminating
and sometimes fully adequate analytical theory for analyzing
composite plasmonic structures~\cite{Prodan:2003}. For example,
the plasmonic resonances of a nanoparticle dimer result from the
hybridization of the plasmons of its constituent NPs. In order of
increasing frequency, we have the bonding-dipolar plasmon (BDP),
bonding-quadrupolar plasmon (BQP), and higher-order plasmon
modes~\cite{romero_oex14}.
When the nanoparticles overlap, the charge-transfer plasmon (CTP)
emerges~\cite{zuloaga_nl9}. While the evolution of plasmonic
resonances in well-separated dimers is well understood through the
hybridization model and classical computational electrodynamics,
the situation becomes more complex at nanoscale gap sizes.
Non-classical effects, including nonlocality, electron spillover,
and nonlocal damping, emerge, as evidenced by both experimental and
computational studies~\cite{Mortensen:2021,Cirac2012Probing,Savage2012,Dionne2012QuantumPlasmon,Dionne2013,Khurgin2017LandauDamping}.
However, the existing understanding originates almost exclusively
from a spectroscopic perspective. As a result, in the transition regime,
information about the linewidth, quality factor, and even resonance
position of the plasmons cannot be obtained without ambiguity,
particularly due to nonlocal damping~\cite{tserkezis_prb96}.
Resolving this intricate issue requires a rigorous mode theory that
can simultaneously account for non-classical effects.

In this work, we conduct a systematic modal analysis of a plasmonic
nanosphere dimer using a quantum hydrodynamic-based quasinormal mode
(QNM) theory. We rigorously trace the entire evolution of the dominant
plasmon modes, bridging the well-separated and overlapping limits, and
identifying critical gap sizes. The analysis accurately reveals how key
modal properties, such as the complex eigenfrequency and quality factor,
respond to the narrowing gap. For the CTP, we introduce an index to
characterize the efficiency of charge transfer across the gap. Moreover,
the modal analysis enables a quantitative assessment of the significance
of nonlocal damping in a mode-resolved manner.

\section{Methodology}
Before delving into the results, we elaborate on the theoretical method
necessary for systematically studying plasmonic gap structures with gap
sizes ranging from positive to negative values. Considering the relevant
deca-nanometer particle size and the need for compatibility with QNM theory,
we must adopt a semiclassical approach rather than relying on first-principles
methods. Among the popular semiclassical models~\cite{Yang2019NatureFramework,Yan2015,Christensen_prl2017,Raza2011},
self-consistent QHT stands out for its ability to describe all relevant
non-classical effects, particularly the detailed stationary electron
distribution in the gap region~\cite{Toscano:2015,Yan_prb2015,Ciraci:2016,Ciraci:2017,Baghramyan:2021}.
Regarding the QNM theory~\cite{Lalanne2018LigntInteraction,YanWei2018PRB},
several extensions have been developed beyond the classical
formulation~\cite{Zhou2022QSR,Zhou:2021,Stephen2017NonlocalQNM,Binkowski2019ModalAnalysis}.
In particular, one of us recently reported the generalized Lorentz model (GLM)~\cite{Zhou:2021}, which enables the establishment of a QNM theory
based on any continuum response models, including QHT. Therefore,
we deploy QHT and the corresponding QNM theory in this work.

In QHT, the conduction electrons responsible for the plasmonic response
of a metallic structure are characterized by the electron density
$n (\mathbf{r}, t)$ and the velocity field $\mathbf{v} (\mathbf{r}, t)$.
Restricting to the linear response under weak light excitation, the
plasmon is built perturbatively on the stationary distribution
$n_{0} (\mathbf{r})$. Here, $n_{0} (\mathbf{r})$ is self-consistently
determined by solving the force equilibrium equation
\begin{equation}
\nabla \left( \frac{\delta G}{\delta n} \right)_{0} -
q_{\mathrm{e}} \mathbf{E}_{0} = 0
\label{Eq:n0}
\end{equation}
together with Poisson's equation $\nabla^{2} \phi_{0} =
(n_{+} - n_{0}) q_{\mathrm{e}}/ \varepsilon_{0}$. The subscript
“0” denotes the stationary components. The electrostatic field is
related with the static potential as $\mathbf{E}_{0} = -\nabla \phi_{0}$.
In eq~\ref{Eq:n0}, the non-classical force derives from the internal
energy functional $G = \int d\mathbf{r} g[n (\mathbf{r}, t)]$. The energy
density $g[n (\mathbf{r}, t)] = t_{\mathrm{TF}} + t_{\mathrm{W}} +
e_{\mathrm{XC}}$ consists of the Thomas--Fermi kinetic energy
$t_{\mathrm{TF}} = \frac{\hbar^{2}\pi^{3}}{10 m_\mathrm{e}}
\left( \frac{3n}{\pi}\right)^{5/3}$, the von~Weizs\"{a}cker kinetic
energy $t_{\mathrm{W}} = \frac{\lambda_{\mathrm{W}} \hbar^{2}}{8 m_{\mathrm{e}}} \frac{\nabla n \cdot \nabla n}{n}$, and the exchange-correlation energy
$e_{\mathrm{XC}} =  -\left(0.0588  +
 \frac{0.035 }{0.6204 + 7.8\times a_{\mathrm{H}} n^{1/3}}\right)\frac{q_{\mathrm{e}}^{2}}{\varepsilon_{0}}n^{4/3}$.
In the above, $\varepsilon_{0}$, $\hbar$, $a_{\mathrm{H}} = 0.529$\,{\AA},
$q_{\mathrm{e}} < 0$, $m_{\mathrm{e}}$, $\lambda_{\mathrm{W}} = 0.12$,
and $n_{+}$ denote respectively the vacuum permittivity, the reduced Planck
constant, the Bohr radius, the electron charge, the electron mass, the
von~Weizs\"{a}cker parameter, and the uniform number density for the positive
ionic lattice according to the jellium approximation. For the
exchange-correlation energy, Wigner's local-density approximation is
employed. The $n_{0}$ distribution, incorporating the above non-classical
contributions, can accurately describe electron spillover both at the
surface and in the gap region.

On top of the stationary electron distribution, the linear plasmonic response
follows from
\begin{equation}
(-i \omega + \gamma) \mathbf{J} = 
\frac{q_{\mathrm{e}}}{m_{\mathrm{e}}} 
\left[n_{0} q_{\mathrm{e}} \mathbf{E} - 
n_{0} \nabla \left (\frac{\delta G}{\delta n}\right)_{1} + 
\nabla \cdot\bar{\bar{\sigma}}\right]
\label{Eq:Lin_Resp}
\end{equation}
in terms of the induced electron density $n_{1} (\mathbf{r}, t)$ and the
current density $\mathbf{J} (\mathbf{r}, t) = n_{0} (\mathbf{r}) q_{\mathrm{e}} \mathbf{v} (\mathbf{r}, t)$.
The linear-order terms are indexed with a subscript ``1'' in cases of
ambiguity, and we have already suppressed the factor $e^{-i\omega t}$
according to the assumed time-harmonic convention. Here, the non-classical
effects manifest themselves in the plasmonic response through the linear-order
components of $\frac{\delta G}{\delta n}$. Apart from the phenomenological
damping with the rate of $\gamma$, a non-adiabatic energy functional in QHT
serves as an effective avenue to model the nonlocal damping. A current-dependent
functional proves to be sufficiently effective for the current purpose and
appears as the viscous stress tensor $\bar{\bar{\sigma}}$ in eq~\ref{Eq:Lin_Resp}~\cite{Ciraci:2017}.
The Cartesian components of $\bar{\bar{\sigma}}$ are given by $\sigma_{\mu \nu} =
\eta_{\mathrm{k}} ( \frac{\partial v_{\mu}}{\partial x_{\nu}} + 
\frac{\partial v_{\nu}}{\partial x_{\mu}} - 
\frac{2}{3} \delta_{\mu\nu} \nabla \cdot \mathbf{v} )$ with the coefficient
$\eta_{\mathrm{k}} = 14 \hbar \left(60 r_{\mathrm{s0}}^{-3/2} + 
80 r_{\mathrm{s0}}^{-1} - 40 r_{\mathrm{s0}}^{-2/3} + 
62 r_{\mathrm{s0}}^{-1/3}\right)^{-1} n_{0}$ and the position-dependent
Wigner--Seitz radius $r_{\mathrm{s0}} = \sqrt[3]{3/(4\pi a_{\mathrm{H}}^{3} n_{0}})$.
By coupling with Maxwell's equations and the continuity equation, $-i \omega
q_{\mathrm{e}} n_{1} + \nabla \cdot \mathbf{J} = 0$, eq~\ref{Eq:Lin_Resp} forms
the closed linear-response theory. In studying the transition of the plasmonic
response, the inhomogeneous $n_{0}$ distribution and nonlocal damping play a
significant role. To underscore their impact, we also compare with a hydrodynamic
theory (HT) that assumes a homogeneous $n_{0} = n_{+}$ and neglects nonlocal
damping. The linear response according to this HT is typically governed by~\cite{Raza2011,raza_JPCM_2015}
\begin{equation}
\beta^{2} \nabla (\nabla \cdot \mathbf{J} ) +
\omega (\omega + i\gamma) \mathbf{J} =
i \omega \omega_{\mathrm{p}}^{2} \varepsilon_{0} \mathbf{E}
.
\label{Eq:HT}
\end{equation}
Here, $\omega_{\mathrm{p}}^{2} = q_{\mathrm{e}}^{2}
n_{+}/(m_{\mathrm{e}} \varepsilon_{0})$ is the plasma frequency.
The parameter $\beta$ can be understood as a characteristic velocity,
arising from the Thomas--Fermi energy~\cite{halevi_prb51,Wegner:2023}.
As an energy functional of $n_0$, the local-density approximation of exchange-correlation similarly contributes to this parameter.
For the Wigner exchange-correlation approximation and the metal sodium (Na) concerned in this work, we find the parameter takes the value $\beta = \sqrt{12/125} v_{\mathrm{F}}$.

In addition to the direct calculation of the linear response, a QHT-based
QNM theory is required to conduct modal analysis. As a continuum theory,
QHT fits into the GLM. For the linear response, eq~\ref{Eq:Lin_Resp} can
be reformulated as~\cite{Zhou:2021}
\begin{equation} 
\omega^{2} \mathbf{P} + i\omega \rho \hat{\Gamma}\mathbf{P} -
\rho \hat{\Theta} \mathbf{P} + 
\varepsilon_{0} \omega_{\mathrm{p}}^{2} \rho\mathbf{E} = 0
,
 \label{Eq:GLM}
\end{equation}
where $\mathbf{P}$ and $\rho = n_{0}/n_{+}$ are the polarization field and
the normalized stationary electron density, respectively. The nonlocal damping
and restoring force operators of the QHT turn out to be
\begin{subequations}
    \begin{equation}
[\hat{\Gamma}_{1}]_{jj'} = 
\tfrac{\tfrac{2}{3} \delta_{ij} \delta_{i'j'} - 
\delta_{i'j} \delta_{ij'} - \delta_{i'i}\delta_{jj'}}
{m_{\mathrm{e}} n_{+}}
\frac{\partial_{i}}{\rho}
\left[ \eta_{\mathrm{k}} \partial_{i'} 
\left( \frac{1}{\rho} \cdot \right)\right]
,
\label{Eq:Gamma}
\end{equation}
\begin{equation}
\hat{\Theta} =
\nabla K_{1} (\nabla \cdot )-
\nabla \nabla \cdot K_{2} \nabla( \nabla\cdot)
,
\label{Eq:Theta}
\end{equation}
\end{subequations}
where $K_{1}$ and $K_{2}$ are both functions of $\rho$~\cite{Zhou:2021}.
The full damping-force operator is given by $\hat{\Gamma} =
\hat{\Gamma}_{1} + \gamma/\rho$. Written in the GLM form, eq~\ref{Eq:Lin_Resp}
conveniently converts into a linear eigenvalue problem with the eigenvalue
$\omega$ and eigenvector $[\mathbf{E}, \mathbf{H}, \mathbf{P},\mathbf{J}]$.
The latter defines the QNM in QHT. More importantly, the GLM formulation also
provides a framework upon which the theorems for local-response electromagnetic
media can be generalized to non-classical ones. From the generalized Poynting
theorem, we identify the absorption power to be
\begin{equation}
\mathbf{P}_{\mathrm{abs}} =
\frac{1}{\varepsilon_{0} \omega_{\mathrm{p}}^{2}}
\int_{\Omega} d\mathbf{r} \mathbf{J} \cdot
\left(\hat{\Gamma}_{1}^{*} + 
\frac{\gamma}{\rho} \right) \mathbf{J}^{*}
.
\label{Eq:Abs}
\end{equation}
The absorption consists of two components: $\mathbf{P}_\mathrm{abs,NL}$
due to nonlocal damping and the phenomenological part
$\mathbf{P}_{\mathrm{abs}, 0}$. The HT, as a simplified version of the QHT,
is essentially a specialized form of the GLM, with the force operators
given by $\hat{\Theta} = -\nabla \beta^{2} (\nabla \cdot)$ and
$\hat{\Gamma} = \gamma$. The QNM theory for HT then follows from the GLM
framework.

\section{Results and discussion}

\begin{figure}[htbp]
\centering
\includegraphics[width=.6\textwidth]{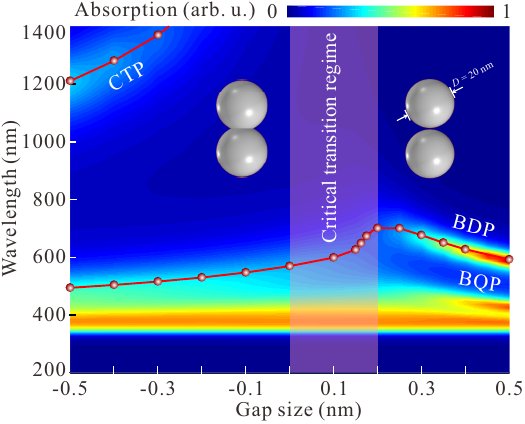}
\caption{Colormap of the absorption spectrum as a function
of wavelength $\lambda$ and gap size $g$ for an incident plane wave
polarized along the axis of a $D = 20$\,nm sodium nanosphere
dimer (see insets). The bonding dipolar plasmon (BDP) and
charge-transfer plasmon (CTP) modes are traced (red lines), and superimposed on those data points obtained from QNM analysis.
}
\label{Fig:Schematics}
\end{figure}

Prevalent in plasmonics, a nanogap can be constructed in various plasmonic
structures, including, \emph{e.g.}, nanoparticle dimers~\cite{yoon_acsphot6},
particle assemblies~\cite{tserkezis_part31}, and particle-on-mirror
configurations~\cite{Baumberg2019ExtremeNanophotonics}. Without loss of generality, in
this work we use a nanosphere dimer as an example of typical plasmonic gap
structures. As sketched in the insets of Figure~\ref{Fig:Schematics}, the
dimer consists of two sodium nanospheres of diameter $D = 20$\,nm, with
the gap size $g$ varying from positive to negative, the latter corresponding
to overlapping spheres. The material is assumed to be the simple alkali
metal sodium (Na) in order to relieve our study from
the influence of the bound electrons typically contributing to the optical
response of noble metals. Sodium has a bulk electron density $n_{+} =
2.52 \times 10^{28}$\,m$^{-3}$ and phenomenological damping rate
$\gamma = 0.066$\,eV. To set the stage for the discussion, we first examine
the plasmonic response of the dimer from the spectroscopic perspective.
We assume a plane wave polarized along the dimer axis incident on the dimer
to excite the LSP modes of interest. Solving eq~\ref{Eq:Lin_Resp}, we
obtain the absorption spectra of the dimer for $g \in [-0.5, 0.5]$\,nm, as
rendered into the colormap in Figure~\ref{Fig:Schematics}. We designate the
shaded gap size region with 
$g \in [0, 0.2]$\,nm 
as the critical transition
regime (CTR). When $g > 0.2$\,nm, the plasmons of the individual NPs are
capacitively coupled, with the coupling strength increasing as the gap size
decreases. Due to strong hybridization, both the BDP and BQP redshift,
moving out of the major resonance formed by the coalescing higher-order
plasmons. For the overlapped dimer, while the CTP beyond 1\,$\upmu$m is
clearly observable, another plasmon manifests vaguely as a shoulder around the 500\,nm wavelength range.
The identity of this latter plasmon
cannot be easily discerned from the spectrum, and it is sometimes assigned
as the second charge-transfer plasmon (CTP') in the 
literature~\cite{tserkezis_oex22,Lei2018}.
The difficulty arises partly due to the significantly broadened and merged
resonance caused by nonlocal damping. In the CTR, nonlocal damping, along with electron spillover, can make the spectra nearly featureless. As a result, it
is intricate to unravel how the plasmonic resonance properties of the dimer
are correlated between the separated and overlapped regimes. We focus on the
fundamental CTP and BDP, and superimpose their eigenwavelengths (real parts)
onto the colormap in Figure~\ref{Fig:Schematics}. Outside the CTR, the 
eigenwavelength data from the QNM theory clearly overlap with the resonance
peak or shoulder positions in the spectra. Inside the CTR, the eigenwavelength
data unequivocally disclose that the BDP continuously evolves across the CTR.
With the superimposed data points, the imaginary parts of the eigenwavelengths
are discarded. A complete picture is obtained when the full information from
the QNMs is considered.

\begin{figure}[htbp]
\centering
\includegraphics[width=.6\textwidth]{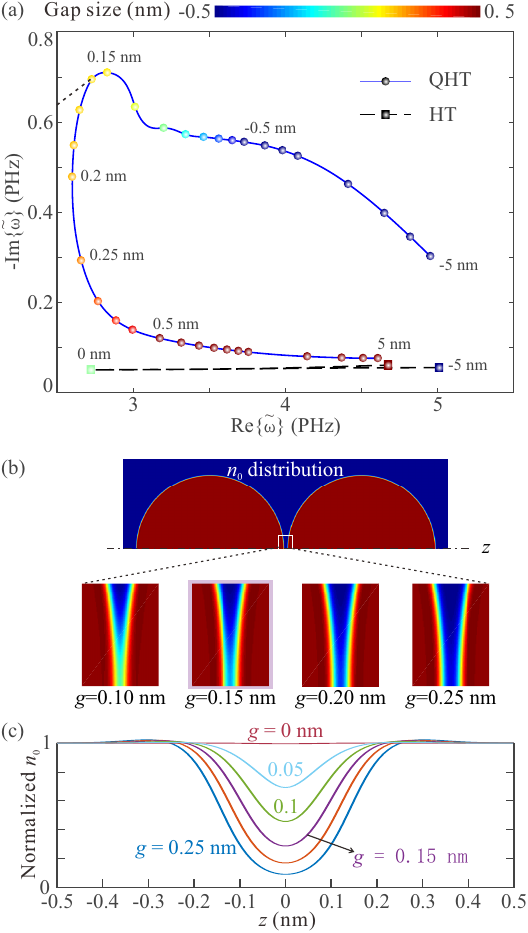}
\caption{Bonding dipolar plasmon mode evolution.
(a) The eigenfrequencies of the BDP mode according to the quantum
hydrodynamic (solid lines, circles) and hard-wall hydrodynamic
(long dashed lines, squares) theories on the complex frequency
plane. The color of the markers denotes the gap size. The thin
dashed line on the top-left corner points to the origin of the coordinate and marks the lowest-quality factor region.
(b) The stationary electron density $n_{0}$ distributions in the gap region
when the gap size crosses the threshold for quantum tunneling.
(c) The $n_{0}$ distributions on the dimer axis for a series of
positive gap sizes.}
\label{Fig:BDP}
\end{figure}

We begin the QNM analysis with the BDP, which typically contributes the
strongest plasmonic enhancement. Applying the QHT-based QNM theory, we
find the BDP mode of the dimer across an extended range of gap sizes,
$g\in [-5, 5]$\,nm. The evolution of the BDP mode is visualized on the complex
frequency plane in Figure~\ref{Fig:BDP}a, with the gap size indicated by
the color of the circle markers. Note that tracing on the complex wavelength
plane is another viable option for visualizing the mode evolution. We opt
for the present choice because both the real and the imaginary parts of a
complex frequency have clear physical meanings, considering the time-harmonic
factor 
$e^{-i(\omega_{\mathrm{R}} - i|\omega_{\mathrm{I}}|)t}$
used to describe
an optical field in the frequency domain. With the gap size spanning both the
well-separated and the overlapped cases, the eigenfrequency trace demonstrates
the entire evolution. Both the real and imaginary parts exhibit a back-bending
feature. The resonance frequency initially redshifts as the capacitive coupling
augments with the gap size approaching the CTR from the positive side. Then,
as the gap size decreases further, the coupling is diminished by nonlocality
and electron spill-out, causing the BDP to blueshift. In contrast, the similar
back-bending behavior of the imaginary part results from the maximal nonlocal
damping occurring in the CTR. To emphasize the impact of electron spill-out
and nonlocal damping, we also present in Figure~\ref{Fig:BDP}a the eigenfrequency
trace of the BDP calculated using HT, indicated by the square markers.
The evolution according to HT correctly reflects the rise and fall of the
coupling strength as electron nonlocality is considered. However, the
much-simplified overall pattern and the nearly constant imaginary part
result from the neglect of electron spill-out and nonlocal damping.
At large gap sizes, the two methods agree well; however, there is still a
sizable discrepancy in the imaginary part even when the gap size is
$g = -5$\,nm. The discrepancy arises because nonlocal damping remains
significant for the overlapped dimer, as will be discussed later.

Next, we elaborate on the extrema associated with the back-bending pattern.
The maximum of the real part occurs at $g_{\mathrm{R}} = 0.2$\,nm, which
precisely marks the entry into the CTR. More dramatic transition takes
place inside the CTR. The nonlocal damping and hence the imaginary part
peak at the gap size of $g_{\mathrm{I}} = 0.15$\,nm. The ordering of the
appearance of the extrema lies in the distinct underlying mechanisms.
While the capacitive coupling strength falls once non-classical effects
start to engage, other important aspects only respond to the electron
spill-out effect at a smaller gap. Inside the CTR the field confinement
continues growing stronger with the shrinking gap until the electron
screening by the spill-out puts an end to the growth. Concomitantly,
the maximal nonlocal damping and imaginary part are accompanied by the
emergence of a considerable electron spill-out. These arguments are in
fact supported by the stationary electron distributions obtained with the
self-consistent QHT. As illustrated in Figure~\ref{Fig:BDP}b, the electron
density $n_{0}$ in the gap region is negligible for $g \geq 0.2$\,nm, and
becomes visible at $g \leq 0.15$\,nm. The $n_{0}$ distributions along the
dimer axis are further depicted in Figure~\ref{Fig:BDP}c for a quantitative
description. 30\% of the homogeneous $n_{+}$ appears as the threshold to
trigger the extrema. The imaginary part of the eigenfrequency is related to
another key property of the BDP plasmon mode, \emph{i.e.} the quality factor $Q$.
Defined as $Q = \omega_{\mathrm{R}}/2 |\omega_{\mathrm{I}}|$, it is directly
determined by the slope of the line connecting the origin and a point on the
trace. The lowest quality factor obviously occurs at the point where the
line directing to the origin coincides with the tangent of the trace. In Figure~\ref{Fig:BDP}a,
the lowest quality factor point is pictorially pinpointed with the dashed
tangent line. The corresponding gap size is $0.16$\,nm, only slightly larger
than $g_{\mathrm{I}}$ which we deem as a more critical gap size during the
transition.

\begin{figure}[htbp]
\centering
\includegraphics[width=1\textwidth]{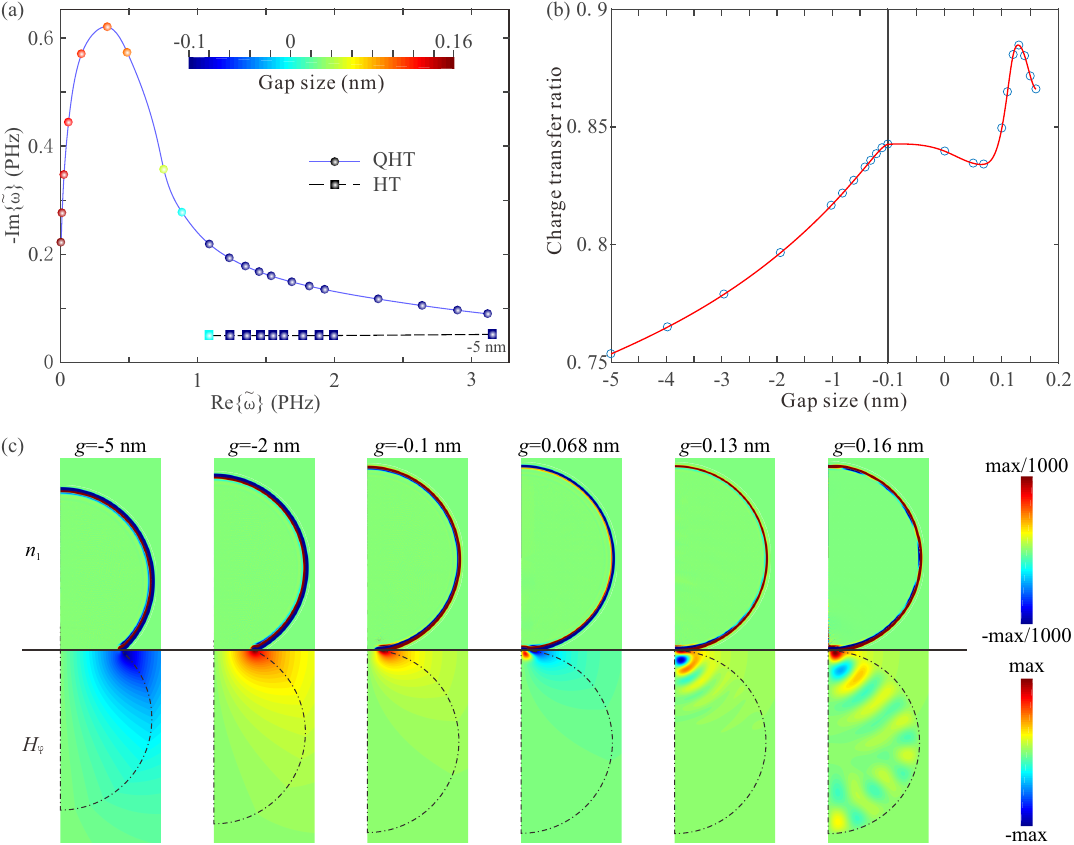}
\caption{CTP mode evolution.
(a) The eigenfrequencies of the CTP mode according to QHT (solid lines)
and HT (dashed line) on the complex frequency plane. The color of the
markers denotes the gap size.
(b) The charge-transfer ratio (defined in eq~\ref{Eq:ratio_CT}) of the
CTP as a function of the gap size. The abscissa from $-0.1$\,nm to $0.2$\,nm
is zoomed-in to resolve the variations.
(c) The modal distributions of the induced electron density
$\widetilde{n}_{1}$ (upper panels) and magnetic field
$\widetilde{H}_{\phi}$ (lower panels) for the CTPs at a series of
representative gap sizes.
}
\label{Fig:CTP}
\end{figure}

The CTP appears as the sole resonance in the long
wavelength range as witnessed at the top-left corner of
Figure~\ref{Fig:Schematics}. It is of particular interest owing to its
sensitivity to the conductance of the nanogap and as a unique resonance
extendable to the terahertz range. We proceed to the QNM analysis of the
CTP in this section. By employing the QHT based QNM theory, we trace the
CTP mode for the gap sizes from $-5$\,nm until it practically ceases to
exist. The resulting trace is presented in Figure~\ref{Fig:CTP}a. Since
the CTP is well defined for the overlapped dimer, we embark on the
discussion starting from the negative gap sizes. The real part of
the eigenfrequency exhibits a monotonous redshift with increasing gap
size. The rate of redshifting is, however, remarkably accelerated after
entering the positive gap size range. The observation can be interpreted
by associating the CTP resonance frequency with the time required to
transfer the charge between the NPs and thus with the conductance of
the nanogap~\cite{Aizpurua2010}. During the transition from the overlapped
to kissing dimer, the two NPs essentially remain connected with a
conducting neck, so that the conductance decreases slowly and the
redshift is moderate. After entering the CTR, the conducting neck
quickly disappears with the opening of the gap. The conductance drops
abruptly, causing the resonance frequency of the CTP to plunge by more
than two orders of magnitude. On the other hand, the imaginary part
pertinent to dissipation has a maximum in the transition process.
This happens because the CTP suffers a maximal nonlocal damping in
the CTR as does the BDP. Interestingly, meanwhile the resistance hikes
due to the thinning of electron density in the opening gap, the
dissipation becomes alleviated for the further separated dimer.
Nevertheless, the quality factor of the CTP mode is decreasing all
the way as given by the inverse of the slope. In the end, the CTP
ceases to exist as the linewidth relative to the resonance frequency
diverges. This is consistent with the fact that the CTP does not have
a counterpart for the individual NPs and cannot be properly understood
with the hybridization theory. Without the electron spillover, HT
cannot provide a full description of the CTP. The HT-based QNM theory
only describes the CTP mode for the overlapped dimer, as illustrated in
Figure~\ref{Fig:CTP}a. The redshift of the resonance is predicted with
a good accuracy, but the imaginary part again changes little as only
the constant phenomenological damping is included in the HT. The
nonlocal damping as indicated by the discrepancy with the QHT is yet
not as severe as for the BDP.

Charge transfer between the constituent NPs is the most prominent feature
of the CTP. However, there still lacks an index to evaluate the performance
of the CTP for the task. Moreover, in a general sense, other plasmons like
the BDP are also referred to as CTPs, as long as they persist for the
overlapped dimer. It is not clear whether these CTPs are as capable as the
fundamental CTP of transferring electrons. To this end we propose the charge-transfer ratio defined as below
\begin{equation}
r_{\mathrm{CT}} = 
\left| \int_{\mathrm{NP1}} \widetilde{n}_{1} dV \right| 
{\Big/} \int_{\mathrm{NP1}} |\widetilde{n}_{1}| dV
,
\label{Eq:ratio_CT}
\end{equation}
where $\widetilde{n}_{1}$ stands for the induced electron density of
the plasmonic eigenmode, and the volume integrals are performed for one nanoparticle (NP1). The ratio is intended to characterize the proportion
of the induced electrons displaced from the other nanoparticle (NP2) in
the total induced electrons in NP1. Once some local polarization occurs
within NP1, the corresponding contribution to $\widetilde{n}_1$ cancels
in the numerator but not in the denominator, so that $r_{\mathrm{CT}}$
becomes smaller than unity. Only when all the induced electrons come from
NP2, does $r_{\mathrm{CT}}$ reach the maximum or unity. As such, the charge-transfer ratio
$r_{\mathrm{CT}}$ assesses the purity of a plasmon as a CTP. We note that
this ratio only offers the delineation of one aspect of a CTP. Other aspects
like the strength of the oscillation are needed for a complete description.
We specifically evaluate $r_{\mathrm{CT}}$ for the CTP and depict it as a
function of the gap size in Figure~\ref{Fig:CTP}b. The results undoubtedly
prove the CTP indeed boasts a pretty high $r_{\mathrm{CT}}$, generally
above 75\%. In contrast, we find that the value for the BDP stays below
20\% for the overlapped dimer. We focus on the ratio for the CTP, which
displays a nontrivial variation with the gap size. The charge-transfer ratio $r_{\mathrm{CT}}$
continuously grows to about 85\%, until the gap size closes in on the CTR;
then a dip and a peak occur within the CTR. We try to interpret the
nontrivial variation with the help of the modal profiles. The induced
electron density and magnetic field patterns are showcased respectively
in the upper and lower panels of Figure~\ref{Fig:CTP}c for a series of
representative gap sizes. For all the cases, $n_{1}$ concentrates near the
NP surface. More interestingly, multilayers of alternate signs form along
the radial direction, indicating that the CTP assumes a multipole plasmon
feature. Such a multipole plasmon feature apparently limits $r_{\mathrm{CT}}$
to a lower value. In fact, the initial rise of $r_{\mathrm{CT}}$ with the
gap size is observed to pertain to the gradual weakening of the multipolar
pattern. Inside the CTR, $r_{\mathrm{CT}}$ declines slightly upon the
breakup of the conducting neck. Afterwards, $r_{\mathrm{CT}}$ goes on to
rise and culminates at the $0.13$\,nm gap, when the multipolar pattern nearly
disappears. From this point on, the imaginary part of the eigenfrequency
starts to dominate over the real part, and the modal profiles or the responses
at the eigenfrequency may differ from the responses to a real-frequency
excitation. For instance, the latter two modal magnetic fields in
Figure~\ref{Fig:CTP}c propagate with a much shorter wavelength across
the NP, rather than decay monotonously inside the metal as in the former
cases. The interference pattern then induces extra polarization at the
surface, causing $r_{\mathrm{CT}}$ to decline. We remark, in passing, that
$r_{\mathrm{CT}}$ can be calculated for a response to a real-frequency
excitation and maintains the high level for the CTP at real frequency.

\begin{figure}[htbp]
\centering
\includegraphics[width=.55\textwidth]{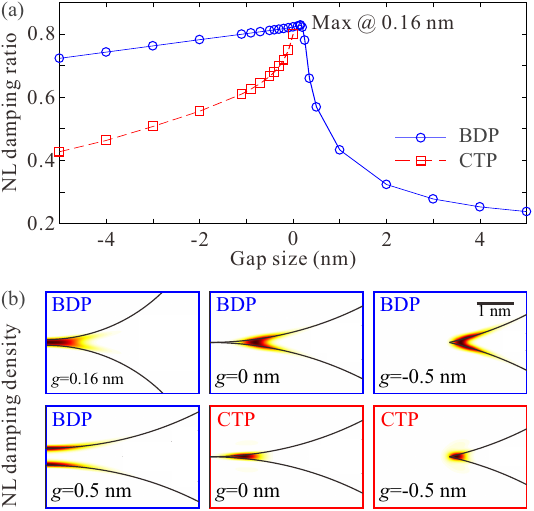}
\caption{Nonlocal damping contribution to the absorption of the sodium dimer.
(a) The nonlocal damping ratio (defined in eq~\ref{Eq:ratio_NLdamping}) as functions of the gap size for both BDP (blue line, circles) and CTP (red line,
squares).
(b) The power density of the nonlocal damping in the gap region for
the representative cases denoted in the label of each panel.
}
\label{Fig:NL_damping}
\end{figure}

As we have demonstrated, nonlocal damping plays an essential role in the
plasmon evolution. But these observations are based so far on qualitative
and comparative analysis. In the last section we take advantage of the GLM
formulation of QHT to quantify the contribution of nonlocal damping. To be
specific, eq~\ref{Eq:Abs} allows to explicitly separate the total absorption
into nonlocal and phenomenological contributions. We therefore introduce the
nonlocal ratio
\begin{equation}
r_{\mathrm{NL}} = 
P_{\mathrm{abs,NL}} / P_{\mathrm{abs}}
\label{Eq:ratio_NLdamping}
\end{equation}
to quantitatively characterize the contribution of nonlocal damping.
Additionally, the expression of the absorption density facilitates the
visualization of the spatial distribution of the nonlocal damping.
We compute $r_{\mathrm{NL}}$ for both the BDP and CTP. Since the
dissipation at a complex frequency is not easily defined, the ratio
is evaluated at the real part of the eigenfrequencies. We plot the
results against the gap size in Figure~\ref{Fig:NL_damping}a. As
expected, the $r_{\mathrm{NL}}$ traces maximize with
$r_{\mathrm{NL}} > 80\%$ in the CTR, and decay for both the overlapped
and separated dimers. For the BDP, the maximal $r_{\mathrm{NL}}$ occurs
at the same gap size when the quality factor minimizes. The nonlocal
damping is strongly enhanced in the gap region, as shown in
Figure~\ref{Fig:NL_damping}b. Outside the CTR, $r_{\mathrm{NL}}$ decays
much faster at the positive gaps than for the overlapped dimer. This is
consistent with the the nonlocal damping distributions given in 
Figure~\ref{Fig:NL_damping}b. While only feeble nonlocal damping remains
at the opposite surfaces of the gap, it is squeezed at the sharp corner
of the neck in the overlapped dimer. The ratio is even still above
75\% at the -5\,nm gap. The sizable nonlocal damping explains the large
discrepancy in the imaginary part of the eigenfrequency computed by QHT
and HT. In comparison, nonlocal damping decays to about 40\% at the $-5$\,nm
gap for the CTP. We can tell the difference from the nonlocal damping
distributions in Figure~\ref{Fig:NL_damping}b. The nonlocal damping density is squeezed
into a smaller region in the corner for the CTP than the BDP, resulting
in relatively weak damping.

\section{Conclusion}
We analyzed mode-transition processes in plasmonic nanogaps, and explored the roles of electron spill-out and nonlocal damping.
Employing a recently developed QNM theory in the QHT framework, we were able to self-consistently follow the evolution of both the real and the imaginary parts of the eigenfrequencies of a sodium nanosphere dimer, shedding light on the distinct mechanisms behind their nontrivial evolution for both the BDP and the main CTP.
In particular, the imaginary part quantitatively reveals how the key properties associated with nonlocal damping, including the linewidth and quality factor, 
evolve in the transition regime.
The vanishing quality factor of the CTP mode with the opening gap helps rationalize its gradual disappearance in the separated dimer.
We further introduced a charge-transfer index that allows to clearly demonstrate the CTP's prominent efficiency of transporting electrons between the nanoparticles.
Lastly we quantified the contributions of different origins to plasmonic damping, unveiling the dominance of nonlocal damping in the nearly touching and overlapped situations.
While focused on nanosphere dimers, our conclusions generally apply to any other gap structure, and provide a deeper understanding of the physics governing plasmonic nanogaps of {\AA}ngstr{\"o}m size.

\begin{acknowledgement}

The Center for Polariton-driven Light--Matter Interactions (POLIMA) is sponsored by the Danish National Research Foundation (Project No.~DNRF165). P.~Z. acknowledges financial support from the National Natural Science Foundation of China (Grant No. 12274160).

\end{acknowledgement}

\bibliography{achemso-demo}

\providecommand{\latin}[1]{#1}
\makeatletter
\providecommand{\doi}
  {\begingroup\let\do\@makeother\dospecials
  \catcode`\{=1 \catcode`\}=2 \doi@aux}
\providecommand{\doi@aux}[1]{\endgroup\texttt{#1}}
\makeatother
\providecommand*\mcitethebibliography{\thebibliography}
\csname @ifundefined\endcsname{endmcitethebibliography}  {\let\endmcitethebibliography\endthebibliography}{}
\begin{mcitethebibliography}{72}
\providecommand*\natexlab[1]{#1}
\providecommand*\mciteSetBstSublistMode[1]{}
\providecommand*\mciteSetBstMaxWidthForm[2]{}
\providecommand*\mciteBstWouldAddEndPuncttrue
  {\def\EndOfBibitem{\unskip.}}
\providecommand*\mciteBstWouldAddEndPunctfalse
  {\let\EndOfBibitem\relax}
\providecommand*\mciteSetBstMidEndSepPunct[3]{}
\providecommand*\mciteSetBstSublistLabelBeginEnd[3]{}
\providecommand*\EndOfBibitem{}
\mciteSetBstSublistMode{f}
\mciteSetBstMaxWidthForm{subitem}{(\alph{mcitesubitemcount})}
\mciteSetBstSublistLabelBeginEnd
  {\mcitemaxwidthsubitemform\space}
  {\relax}
  {\relax}

\bibitem[Schuller \latin{et~al.}(2010)Schuller, Barnard, Cai, Jun, White, and Brongersma]{Brongersma2010Plasmonics}
Schuller,~J.~A.; Barnard,~E.~S.; Cai,~W.; Jun,~Y.~C.; White,~J.~S.; Brongersma,~M.~L. Plasmonics for extreme light concentration and manipulation. \emph{Nat. Mater.} \textbf{2010}, \emph{9}, 193--204\relax
\mciteBstWouldAddEndPuncttrue
\mciteSetBstMidEndSepPunct{\mcitedefaultmidpunct}
{\mcitedefaultendpunct}{\mcitedefaultseppunct}\relax
\EndOfBibitem
\bibitem[Halas \latin{et~al.}(2011)Halas, Lal, Chang, Link, and Nordlander]{Halas:2011}
Halas,~N.~J.; Lal,~S.; Chang,~W.-S.; Link,~S.; Nordlander,~P. Plasmons in strongly coupled metallic nanostructures. \emph{Chem. Rev.} \textbf{2011}, \emph{111}, 3913--3961\relax
\mciteBstWouldAddEndPuncttrue
\mciteSetBstMidEndSepPunct{\mcitedefaultmidpunct}
{\mcitedefaultendpunct}{\mcitedefaultseppunct}\relax
\EndOfBibitem
\bibitem[Baumberg \latin{et~al.}(2019)Baumberg, Aizpurua, Mikkelsen, and Smith]{Baumberg2019ExtremeNanophotonics}
Baumberg,~J.~J.; Aizpurua,~J.; Mikkelsen,~M.~H.; Smith,~D.~R. Extreme nanophotonics from ultrathin metallic gaps. \emph{Nat. Mater.} \textbf{2019}, \emph{18}, 668--678\relax
\mciteBstWouldAddEndPuncttrue
\mciteSetBstMidEndSepPunct{\mcitedefaultmidpunct}
{\mcitedefaultendpunct}{\mcitedefaultseppunct}\relax
\EndOfBibitem
\bibitem[Gramotnev and Bozhevolnyi(2010)Gramotnev, and Bozhevolnyi]{Gramotnev:2010}
Gramotnev,~D.~K.; Bozhevolnyi,~S.~I. Plasmonics beyond the diffraction limit. \emph{Nat. Photon.} \textbf{2010}, \emph{4}, 83--91\relax
\mciteBstWouldAddEndPuncttrue
\mciteSetBstMidEndSepPunct{\mcitedefaultmidpunct}
{\mcitedefaultendpunct}{\mcitedefaultseppunct}\relax
\EndOfBibitem
\bibitem[Xu \latin{et~al.}(2000)Xu, Aizpurua, K\"{a}ll, and Apell]{xu_pre62}
Xu,~H.; Aizpurua,~J.; K\"{a}ll,~M.; Apell,~P. Electromagnetic contributions to single-molecule sensitivity in surface-enhanced {Raman} scattering. \emph{Phys. Rev. E} \textbf{2000}, \emph{62}, 4318--4324\relax
\mciteBstWouldAddEndPuncttrue
\mciteSetBstMidEndSepPunct{\mcitedefaultmidpunct}
{\mcitedefaultendpunct}{\mcitedefaultseppunct}\relax
\EndOfBibitem
\bibitem[Acuna \latin{et~al.}(2012)Acuna, M\"{o}ller, Holzmeister, Beater, Lalkens, and Tinnefeld]{acuna_sci338}
Acuna,~G.~P.; M\"{o}ller,~F.~M.; Holzmeister,~P.; Beater,~S.; Lalkens,~B.; Tinnefeld,~P. Fluorescence enhancement at docking sites of {DNA-directed} self-assembled nanoantennas. \emph{Science} \textbf{2012}, \emph{338}, 506--510\relax
\mciteBstWouldAddEndPuncttrue
\mciteSetBstMidEndSepPunct{\mcitedefaultmidpunct}
{\mcitedefaultendpunct}{\mcitedefaultseppunct}\relax
\EndOfBibitem
\bibitem[Yu \latin{et~al.}(2019)Yu, Peng, Yang, and Li]{Yu:2019}
Yu,~H.; Peng,~Y.; Yang,~Y.; Li,~Z.-Y. Plasmon-enhanced light-matter interactions and applications. \emph{npj Comp. Mater.} \textbf{2019}, \emph{5}, 45\relax
\mciteBstWouldAddEndPuncttrue
\mciteSetBstMidEndSepPunct{\mcitedefaultmidpunct}
{\mcitedefaultendpunct}{\mcitedefaultseppunct}\relax
\EndOfBibitem
\bibitem[Gon\c{c}alves \latin{et~al.}(2020)Gon\c{c}alves, Christensen, Rivera, Jauho, Mortensen, and Solja\v{c}i\'{c}]{Goncalves:2020}
Gon\c{c}alves,~P.~A.~D.; Christensen,~T.; Rivera,~N.; Jauho,~A.-P.; Mortensen,~N.~A.; Solja\v{c}i\'{c},~M. Plasmon–emitter interactions at the nanoscale. \emph{Nat. Commun.} \textbf{2020}, \emph{11}, 366\relax
\mciteBstWouldAddEndPuncttrue
\mciteSetBstMidEndSepPunct{\mcitedefaultmidpunct}
{\mcitedefaultendpunct}{\mcitedefaultseppunct}\relax
\EndOfBibitem
\bibitem[Zhang \latin{et~al.}(2017)Zhang, Protsenko, Sandoghdar, and Chen]{Chen2017}
Zhang,~P.; Protsenko,~I.; Sandoghdar,~V.; Chen,~X.-W. A single-emitter gain medium for bright coherent radiation from a plasmonic nanoresonator. \emph{ACS Photonics} \textbf{2017}, \emph{4}, 2738--2744\relax
\mciteBstWouldAddEndPuncttrue
\mciteSetBstMidEndSepPunct{\mcitedefaultmidpunct}
{\mcitedefaultendpunct}{\mcitedefaultseppunct}\relax
\EndOfBibitem
\bibitem[Brongersma \latin{et~al.}(2015)Brongersma, Halas, and Nordlander]{brongersma_natnano10}
Brongersma,~M.~L.; Halas,~N.~J.; Nordlander,~P. Plasmon-induced hot carrier science and technology. \emph{Nat Nanotechnol.} \textbf{2015}, \emph{10}, 25--34\relax
\mciteBstWouldAddEndPuncttrue
\mciteSetBstMidEndSepPunct{\mcitedefaultmidpunct}
{\mcitedefaultendpunct}{\mcitedefaultseppunct}\relax
\EndOfBibitem
\bibitem[Li and Valentine(2017)Li, and Valentine]{li_nanoph6}
Li,~W.; Valentine,~J.~G. Harvesting the loss: surface plasmon-based hot electron photodetection. \emph{Nanophotonics} \textbf{2017}, \emph{6}, 177--191\relax
\mciteBstWouldAddEndPuncttrue
\mciteSetBstMidEndSepPunct{\mcitedefaultmidpunct}
{\mcitedefaultendpunct}{\mcitedefaultseppunct}\relax
\EndOfBibitem
\bibitem[Khurgin(2019)]{khurgin_faraday214}
Khurgin,~J.~B. Hot carriers generated by plasmons: where are they generated and where do they go from there? \emph{Faraday Discuss.} \textbf{2019}, \emph{214}, 35--58\relax
\mciteBstWouldAddEndPuncttrue
\mciteSetBstMidEndSepPunct{\mcitedefaultmidpunct}
{\mcitedefaultendpunct}{\mcitedefaultseppunct}\relax
\EndOfBibitem
\bibitem[Le \latin{et~al.}(2008)Le, Brandl, Urzhumov, Wang, Kundu, Halas, Aizpurua, and Nordlander]{le_nn2}
Le,~F.; Brandl,~D.~W.; Urzhumov,~Y.~A.; Wang,~H.; Kundu,~J.; Halas,~N.~J.; Aizpurua,~J.; Nordlander,~P. Metallic nanoparticle arrays: a common substrate for both surface-enhanced {Raman} scattering and surface-enhanced infrare absorption. \emph{ACS Nano} \textbf{2008}, \emph{2}, 707--718\relax
\mciteBstWouldAddEndPuncttrue
\mciteSetBstMidEndSepPunct{\mcitedefaultmidpunct}
{\mcitedefaultendpunct}{\mcitedefaultseppunct}\relax
\EndOfBibitem
\bibitem[Stockman(2004)]{Stockman2004Nanofocusing}
Stockman,~M.~I. Nanofocusing of optical energy in tapered plasmonic waveguides. \emph{Phys. Rev. Lett.} \textbf{2004}, \emph{93}, 137404\relax
\mciteBstWouldAddEndPuncttrue
\mciteSetBstMidEndSepPunct{\mcitedefaultmidpunct}
{\mcitedefaultendpunct}{\mcitedefaultseppunct}\relax
\EndOfBibitem
\bibitem[Kern \latin{et~al.}(2012)Kern, Gro{\ss}mann, Tarakina, H{\"a}ckel, Emmerling, Kamp, Huang, Biagioni, Prangsma, and Hecht]{Hecht2012AtomicScale}
Kern,~J.; Gro{\ss}mann,~S.; Tarakina,~N.~V.; H{\"a}ckel,~T.; Emmerling,~M.; Kamp,~M.; Huang,~J.-S.; Biagioni,~P.; Prangsma,~J.~C.; Hecht,~B. Atomic-scale confinement of resonant optical fields. \emph{Nano Lett.} \textbf{2012}, \emph{12}, 5504--5509\relax
\mciteBstWouldAddEndPuncttrue
\mciteSetBstMidEndSepPunct{\mcitedefaultmidpunct}
{\mcitedefaultendpunct}{\mcitedefaultseppunct}\relax
\EndOfBibitem
\bibitem[Ding \latin{et~al.}(2016)Ding, Yi, Li, Ren, Wu, Panneerselvam, and Tian]{Ding:2016}
Ding,~S.-Y.; Yi,~J.; Li,~J.-F.; Ren,~B.; Wu,~D.-Y.; Panneerselvam,~R.; Tian,~Z.-Q. Nanostructure-based plasmon-enhanced Raman spectroscopy for surface analysis of materials. \emph{Nat. Rev. Mater.} \textbf{2016}, \emph{1}, 16021\relax
\mciteBstWouldAddEndPuncttrue
\mciteSetBstMidEndSepPunct{\mcitedefaultmidpunct}
{\mcitedefaultendpunct}{\mcitedefaultseppunct}\relax
\EndOfBibitem
\bibitem[Fern{\'a}ndez-Dom{\'i}nguez \latin{et~al.}(2018)Fern{\'a}ndez-Dom{\'i}nguez, Bozhevolnyi, and Mortensen]{Fernandez-Dominguez:2018}
Fern{\'a}ndez-Dom{\'i}nguez,~A.~I.; Bozhevolnyi,~S.~I.; Mortensen,~N.~A. Plasmon-enhanced generation of nonclassical light. \emph{ACS Photonics} \textbf{2018}, \emph{5}, 3447--3451\relax
\mciteBstWouldAddEndPuncttrue
\mciteSetBstMidEndSepPunct{\mcitedefaultmidpunct}
{\mcitedefaultendpunct}{\mcitedefaultseppunct}\relax
\EndOfBibitem
\bibitem[Yuan \latin{et~al.}(2023)Yuan, Bourgeois, Carlin, da~Jornada, and Dionne]{Yuan:2023}
Yuan,~L.; Bourgeois,~B.~B.; Carlin,~C.~C.; da~Jornada,~F.~H.; Dionne,~J.~A. Sustainable chemistry with plasmonic photocatalysts. \emph{Nanophotonics} \textbf{2023}, \emph{12}, 2745--2762\relax
\mciteBstWouldAddEndPuncttrue
\mciteSetBstMidEndSepPunct{\mcitedefaultmidpunct}
{\mcitedefaultendpunct}{\mcitedefaultseppunct}\relax
\EndOfBibitem
\bibitem[Wang \latin{et~al.}(2013)Wang, Abb, Boden, Aizpurua, de~Groot, and Muskens]{Wang2013}
Wang,~Y.; Abb,~M.; Boden,~S.~A.; Aizpurua,~J.; de~Groot,~C.~H.; Muskens,~O.~L. Ultrafast nonlinear control of progressively loaded, single plasmonic nanoantennas fabricated using helium ion milling. \emph{Nano Lett.} \textbf{2013}, \emph{13}, 5647--5653\relax
\mciteBstWouldAddEndPuncttrue
\mciteSetBstMidEndSepPunct{\mcitedefaultmidpunct}
{\mcitedefaultendpunct}{\mcitedefaultseppunct}\relax
\EndOfBibitem
\bibitem[Hu \latin{et~al.}(2016)Hu, Ji, Xu, Yu, Liu, Chen, Chen, Wen, Lifshitz, Wang, Zhang, and Lee]{Lee2016}
Hu,~H.~C.; Ji,~F.; Xu,~Y.; Yu,~J.~Q.; Liu,~Q.~P.; Chen,~L.; Chen,~Q.; Wen,~P.; Lifshitz,~Y.; Wang,~Y.; Zhang,~Q.; Lee,~S.-T. Reversible and precise self-assembly of {Janus} metal-organosilica nanoparticles through a linker-free approach. \emph{ACS Nano} \textbf{2016}, \emph{10}, 7323--7330\relax
\mciteBstWouldAddEndPuncttrue
\mciteSetBstMidEndSepPunct{\mcitedefaultmidpunct}
{\mcitedefaultendpunct}{\mcitedefaultseppunct}\relax
\EndOfBibitem
\bibitem[Manfrinato \latin{et~al.}(2017)Manfrinato, Stein, Zhang, Nam, Yager, Stach, and Black]{Manfrinato2017}
Manfrinato,~V.~R.; Stein,~A.; Zhang,~L.; Nam,~C.-Y.; Yager,~K.~G.; Stach,~E.~A.; Black,~C.~T. Aberration-corrected electron beam lithography at the one nanometer length scale. \emph{Nano Lett.} \textbf{2017}, \emph{17}, 4562--4567\relax
\mciteBstWouldAddEndPuncttrue
\mciteSetBstMidEndSepPunct{\mcitedefaultmidpunct}
{\mcitedefaultendpunct}{\mcitedefaultseppunct}\relax
\EndOfBibitem
\bibitem[Kim \latin{et~al.}(2020)Kim, Mun, Baek, Kim, Hao, Qiu, Jung, and Rho]{Kim2020}
Kim,~I.; Mun,~J.; Baek,~K.~M.; Kim,~M.; Hao,~C.; Qiu,~C.-W.; Jung,~Y.~S.; Rho,~J. Cascade domino lithography for extreme photon squeezing. \emph{Mater. Today} \textbf{2020}, \emph{39}, 89--97\relax
\mciteBstWouldAddEndPuncttrue
\mciteSetBstMidEndSepPunct{\mcitedefaultmidpunct}
{\mcitedefaultendpunct}{\mcitedefaultseppunct}\relax
\EndOfBibitem
\bibitem[Romero \latin{et~al.}(2006)Romero, Aizpurua, Bryant, and Garc\'{i}a~de Abajo]{romero_oex14}
Romero,~I.; Aizpurua,~J.; Bryant,~G.~W.; Garc\'{i}a~de Abajo,~F.~J. Plasmons in nearly touching metallic nanoparticles: singular response in the limit of touching dimers. \emph{Opt. Express} \textbf{2006}, \emph{14}, 9988--9999\relax
\mciteBstWouldAddEndPuncttrue
\mciteSetBstMidEndSepPunct{\mcitedefaultmidpunct}
{\mcitedefaultendpunct}{\mcitedefaultseppunct}\relax
\EndOfBibitem
\bibitem[Zhu \latin{et~al.}(2016)Zhu, Esteban, Borisov, Baumberg, Nordlander, Lezec, Aizpurua, and Crozier]{Crozier2016}
Zhu,~W.; Esteban,~R.; Borisov,~A.~G.; Baumberg,~J.~J.; Nordlander,~P.; Lezec,~H.~J.; Aizpurua,~J.; Crozier,~K.~B. Quantum mechanical effects in plasmonic structures with subnanometre gaps. \emph{Nat. Commun.} \textbf{2016}, \emph{7}, 11495\relax
\mciteBstWouldAddEndPuncttrue
\mciteSetBstMidEndSepPunct{\mcitedefaultmidpunct}
{\mcitedefaultendpunct}{\mcitedefaultseppunct}\relax
\EndOfBibitem
\bibitem[Esteban \latin{et~al.}(2012)Esteban, Borisov, Nordlander, and Aizpurua]{esteban_natcom3}
Esteban,~R.; Borisov,~A.~G.; Nordlander,~P.; Aizpurua,~J. Bridging quantum and classical plasmonics with a quantum-corrected model. \emph{Nat. Commun.} \textbf{2012}, \emph{3}, 825\relax
\mciteBstWouldAddEndPuncttrue
\mciteSetBstMidEndSepPunct{\mcitedefaultmidpunct}
{\mcitedefaultendpunct}{\mcitedefaultseppunct}\relax
\EndOfBibitem
\bibitem[Liu \latin{et~al.}(2019)Liu, Wu, Zhang, Wang, Guo, Su, Zhu, Shao, Chen, Luo, and Lei]{liu_nn13}
Liu,~D.; Wu,~T.; Zhang,~Q.; Wang,~X.; Guo,~X.; Su,~Y.; Zhu,~Y.; Shao,~M.; Chen,~H.; Luo,~Y.; Lei,~D. Probing the in-plane near-field enhancement limit in a plasmonic particle-on-film nanocavity with surface-enhanced {Raman} spectroscopy of graphene. \emph{ACS Nano} \textbf{2019}, \emph{13}, 7644--7654\relax
\mciteBstWouldAddEndPuncttrue
\mciteSetBstMidEndSepPunct{\mcitedefaultmidpunct}
{\mcitedefaultendpunct}{\mcitedefaultseppunct}\relax
\EndOfBibitem
\bibitem[Raza \latin{et~al.}(2015)Raza, Wubs, Bozhevolnyi, and Mortensen]{raza_ol15}
Raza,~S.; Wubs,~M.; Bozhevolnyi,~S.~I.; Mortensen,~N.~A. Nonlocal study of ultimate plasmon hybridization. \emph{Opt. Lett.} \textbf{2015}, \emph{40}, 839--842\relax
\mciteBstWouldAddEndPuncttrue
\mciteSetBstMidEndSepPunct{\mcitedefaultmidpunct}
{\mcitedefaultendpunct}{\mcitedefaultseppunct}\relax
\EndOfBibitem
\bibitem[Yang \latin{et~al.}(2020)Yang, Chen, Ghafoor, Zhang, Zhang, Zhang, Luo, Yang, Sandoghdar, Aizpurua, Dong, and Hou]{Dong2020photoluminescence}
Yang,~B.; Chen,~G.; Ghafoor,~A.; Zhang,~Y.; Zhang,~Y.; Zhang,~Y.; Luo,~Y.; Yang,~J.; Sandoghdar,~V.; Aizpurua,~J.; Dong,~Z.; Hou,~J.~G. Sub-Nanometre Resolution in Single-Molecule Photoluminescence Imaging. \emph{Nat. Photon.} \textbf{2020}, \emph{14}, 693--699\relax
\mciteBstWouldAddEndPuncttrue
\mciteSetBstMidEndSepPunct{\mcitedefaultmidpunct}
{\mcitedefaultendpunct}{\mcitedefaultseppunct}\relax
\EndOfBibitem
\bibitem[Lee \latin{et~al.}(2019)Lee, Crampton, Tallarida, and Apkarian]{Apkarian2019vibrationalmodes}
Lee,~J.; Crampton,~K.~T.; Tallarida,~N.; Apkarian,~V.~A. Visualizing vibrational normal modes of a single molecule with atomically confined light. \emph{Nature} \textbf{2019}, \emph{568}, 78--82\relax
\mciteBstWouldAddEndPuncttrue
\mciteSetBstMidEndSepPunct{\mcitedefaultmidpunct}
{\mcitedefaultendpunct}{\mcitedefaultseppunct}\relax
\EndOfBibitem
\bibitem[Benz \latin{et~al.}(2016)Benz, Schmidt, Dreismann, Chikkaraddy, Zhang, Demetriadou, Carnegie, Ohadi, De~Nijs, Esteban, Aizpurua, and Baumberg]{benz_sci354}
Benz,~F.; Schmidt,~M.~K.; Dreismann,~A.; Chikkaraddy,~R.; Zhang,~Y.; Demetriadou,~A.; Carnegie,~C.; Ohadi,~H.; De~Nijs,~B.; Esteban,~R.; Aizpurua,~J.; Baumberg,~J.~J. Single-molecule optomechanics in ``picocavities''. \emph{Science} \textbf{2016}, \emph{354}, 726--729\relax
\mciteBstWouldAddEndPuncttrue
\mciteSetBstMidEndSepPunct{\mcitedefaultmidpunct}
{\mcitedefaultendpunct}{\mcitedefaultseppunct}\relax
\EndOfBibitem
\bibitem[Li \latin{et~al.}(2021)Li, Zhou, Zhang, and Chen]{Li2021}
Li,~W.; Zhou,~Q.; Zhang,~P.; Chen,~X.-W. Bright optical eigenmode of 1 nm$^3$ mode volume. \emph{Phys. Rev. Lett.} \textbf{2021}, \emph{126}, 257401\relax
\mciteBstWouldAddEndPuncttrue
\mciteSetBstMidEndSepPunct{\mcitedefaultmidpunct}
{\mcitedefaultendpunct}{\mcitedefaultseppunct}\relax
\EndOfBibitem
\bibitem[Wu \latin{et~al.}(2021)Wu, Yan, and Lalanne]{Wu2021}
Wu,~T.; Yan,~W.; Lalanne,~P. Bright Plasmons with Cubic Nanometer Mode Volumes through Mode Hybridization. \emph{ACS Photonics} \textbf{2021}, \emph{8}, 307--314\relax
\mciteBstWouldAddEndPuncttrue
\mciteSetBstMidEndSepPunct{\mcitedefaultmidpunct}
{\mcitedefaultendpunct}{\mcitedefaultseppunct}\relax
\EndOfBibitem
\bibitem[Lu \latin{et~al.}(2024)Lu, Ji, Ye, Zhang, Zhang, and Xu]{Xu2024}
Lu,~Z.; Ji,~J.; Ye,~H.; Zhang,~H.; Zhang,~S.; Xu,~H. Quantifying the Ultimate Limit of Plasmonic Near-field Enhancement. \emph{Nat. Commun.} \textbf{2024}, \emph{15}, 8803\relax
\mciteBstWouldAddEndPuncttrue
\mciteSetBstMidEndSepPunct{\mcitedefaultmidpunct}
{\mcitedefaultendpunct}{\mcitedefaultseppunct}\relax
\EndOfBibitem
\bibitem[Mortensen(2021)]{Mortensen:2021}
Mortensen,~N.~A. Mesoscopic electrodynamics at metal surfaces. \emph{Nanophotonics} \textbf{2021}, \emph{10}, 2563--2616\relax
\mciteBstWouldAddEndPuncttrue
\mciteSetBstMidEndSepPunct{\mcitedefaultmidpunct}
{\mcitedefaultendpunct}{\mcitedefaultseppunct}\relax
\EndOfBibitem
\bibitem[Stamatopoulou and Tserkezis(2022)Stamatopoulou, and Tserkezis]{stamatopoulou_omex12}
Stamatopoulou,~P.~E.; Tserkezis,~C. Finite-size and quantum effects in plasmonics: manifestations and theoretical modelling. \emph{Opt. Mater. Express} \textbf{2022}, \emph{12}, 1869--1893\relax
\mciteBstWouldAddEndPuncttrue
\mciteSetBstMidEndSepPunct{\mcitedefaultmidpunct}
{\mcitedefaultendpunct}{\mcitedefaultseppunct}\relax
\EndOfBibitem
\bibitem[Kern \latin{et~al.}(2015)Kern, Kullock, Prangsma, Emmerling, Kamp, and Hecht]{Hecht2015}
Kern,~J.; Kullock,~R.; Prangsma,~J.; Emmerling,~M.; Kamp,~M.; Hecht,~B. Electrically driven optical antennas. \emph{Nat. Photon.} \textbf{2015}, \emph{9}, 582--586\relax
\mciteBstWouldAddEndPuncttrue
\mciteSetBstMidEndSepPunct{\mcitedefaultmidpunct}
{\mcitedefaultendpunct}{\mcitedefaultseppunct}\relax
\EndOfBibitem
\bibitem[Bigourdan \latin{et~al.}(2016)Bigourdan, Hugonin, Marquier, Sauvan, and Greffet]{Greffet2016}
Bigourdan,~F.; Hugonin,~J.-P.; Marquier,~F.; Sauvan,~C.; Greffet,~J.-J. Nanoantenna for electrical generation of surface plasmon polaritons. \emph{Phys. Rev. Lett.} \textbf{2016}, \emph{116}, 106803\relax
\mciteBstWouldAddEndPuncttrue
\mciteSetBstMidEndSepPunct{\mcitedefaultmidpunct}
{\mcitedefaultendpunct}{\mcitedefaultseppunct}\relax
\EndOfBibitem
\bibitem[Li \latin{et~al.}(2022)Li, Zhou, Zhang, and Chen]{Li2021BiasModulation}
Li,~W.; Zhou,~Q.; Zhang,~P.; Chen,~X.-W. Direct electro plasmonic and optic modulation via a nanoscopic electron reservoir. \emph{Phys. Rev. Lett.} \textbf{2022}, \emph{128}, 217401\relax
\mciteBstWouldAddEndPuncttrue
\mciteSetBstMidEndSepPunct{\mcitedefaultmidpunct}
{\mcitedefaultendpunct}{\mcitedefaultseppunct}\relax
\EndOfBibitem
\bibitem[Zurak \latin{et~al.}(2024)Zurak, Wolff, Meier, Kullock, Mortensen, Hecht, and Feichtner]{Zurak2024}
Zurak,~L.; Wolff,~C.; Meier,~J.; Kullock,~R.; Mortensen,~N.~A.; Hecht,~B.; Feichtner,~T. Modulation of surface response in a single plasmonic nanoresonator. \emph{Sci. Adv.} \textbf{2024}, \emph{10}, eadn5227\relax
\mciteBstWouldAddEndPuncttrue
\mciteSetBstMidEndSepPunct{\mcitedefaultmidpunct}
{\mcitedefaultendpunct}{\mcitedefaultseppunct}\relax
\EndOfBibitem
\bibitem[Marinica \latin{et~al.}(2012)Marinica, Kazansky, Nordlander, Aizpurua, and Borisov]{Borisov2012}
Marinica,~D.; Kazansky,~A.; Nordlander,~P.; Aizpurua,~J.; Borisov,~A.~G. Quantum Plasmonics: Nonlinear Effects in the Field Enhancement of a Plasmonic Nanoparticle Dimer. \emph{Nano Lett.} \textbf{2012}, \emph{12}, 1333--1339\relax
\mciteBstWouldAddEndPuncttrue
\mciteSetBstMidEndSepPunct{\mcitedefaultmidpunct}
{\mcitedefaultendpunct}{\mcitedefaultseppunct}\relax
\EndOfBibitem
\bibitem[Prodan \latin{et~al.}(2003)Prodan, Radloff, Halas, and Nordlander]{Prodan:2003}
Prodan,~E.; Radloff,~C.; Halas,~N.~J.; Nordlander,~P. A hybridization model for the plasmon response of complex nanostructures. \emph{Science} \textbf{2003}, \emph{302}, 419--422\relax
\mciteBstWouldAddEndPuncttrue
\mciteSetBstMidEndSepPunct{\mcitedefaultmidpunct}
{\mcitedefaultendpunct}{\mcitedefaultseppunct}\relax
\EndOfBibitem
\bibitem[Zuloaga \latin{et~al.}(2009)Zuloaga, Prodan, and Nordlander]{zuloaga_nl9}
Zuloaga,~J.; Prodan,~E.; Nordlander,~P. Quantum description of the plasmon resonances of a nanoparticle dimer. \emph{Nano Lett.} \textbf{2009}, \emph{9}, 887--891\relax
\mciteBstWouldAddEndPuncttrue
\mciteSetBstMidEndSepPunct{\mcitedefaultmidpunct}
{\mcitedefaultendpunct}{\mcitedefaultseppunct}\relax
\EndOfBibitem
\bibitem[Cirac{\`\i} \latin{et~al.}(2012)Cirac{\`\i}, Hill, Mock, Urzhumov, Fern{\'a}ndez-Dom{\'\i}nguez, Maier, Pendry, Chilkoti, and Smith]{Cirac2012Probing}
Cirac{\`\i},~C.; Hill,~R.~T.; Mock,~J.~J.; Urzhumov,~Y.; Fern{\'a}ndez-Dom{\'\i}nguez,~A.~I.; Maier,~S.~A.; Pendry,~J.~B.; Chilkoti,~A.; Smith,~D.~R. Probing the ultimate limits of plasmonic enhancement. \emph{Science} \textbf{2012}, \emph{337}, 1072--1074\relax
\mciteBstWouldAddEndPuncttrue
\mciteSetBstMidEndSepPunct{\mcitedefaultmidpunct}
{\mcitedefaultendpunct}{\mcitedefaultseppunct}\relax
\EndOfBibitem
\bibitem[Savage \latin{et~al.}(2012)Savage, Hawkeye, Esteban, Borisov, Aizpurua, and Baumberg]{Savage2012}
Savage,~K.~J.; Hawkeye,~M.~M.; Esteban,~R.; Borisov,~A.~G.; Aizpurua,~J.; Baumberg,~J.~J. Revealing the quantum regime in tunnelling plasmonics. \emph{Nature} \textbf{2012}, \emph{491}, 574--577\relax
\mciteBstWouldAddEndPuncttrue
\mciteSetBstMidEndSepPunct{\mcitedefaultmidpunct}
{\mcitedefaultendpunct}{\mcitedefaultseppunct}\relax
\EndOfBibitem
\bibitem[Scholl \latin{et~al.}(2012)Scholl, Koh, and Dionne]{Dionne2012QuantumPlasmon}
Scholl,~J.~A.; Koh,~A.~L.; Dionne,~J.~A. Quantum plasmon resonances of individual metallic nanoparticles. \emph{Nature} \textbf{2012}, \emph{483}, 421--427\relax
\mciteBstWouldAddEndPuncttrue
\mciteSetBstMidEndSepPunct{\mcitedefaultmidpunct}
{\mcitedefaultendpunct}{\mcitedefaultseppunct}\relax
\EndOfBibitem
\bibitem[Scholl \latin{et~al.}(2013)Scholl, Garc\'ia-Etxarri, Koh, and Dionne]{Dionne2013}
Scholl,~J.~A.; Garc\'ia-Etxarri,~A.; Koh,~A.~L.; Dionne,~J.~A. Observation of quantum tunneling between two plasmonic nanoparticles. \emph{Nano Lett.} \textbf{2013}, \emph{13}, 564--569\relax
\mciteBstWouldAddEndPuncttrue
\mciteSetBstMidEndSepPunct{\mcitedefaultmidpunct}
{\mcitedefaultendpunct}{\mcitedefaultseppunct}\relax
\EndOfBibitem
\bibitem[Khurgin \latin{et~al.}(2017)Khurgin, Tsai, Tsai, and Sun]{Khurgin2017LandauDamping}
Khurgin,~J.; Tsai,~W.-Y.; Tsai,~D.~P.; Sun,~G. Landau damping and limit to field confinement and enhancement in plasmonic dimers. \emph{ACS Photonics} \textbf{2017}, \emph{4}, 2871--2880\relax
\mciteBstWouldAddEndPuncttrue
\mciteSetBstMidEndSepPunct{\mcitedefaultmidpunct}
{\mcitedefaultendpunct}{\mcitedefaultseppunct}\relax
\EndOfBibitem
\bibitem[Tserkezis \latin{et~al.}(2017)Tserkezis, Mortensen, and Wubs]{tserkezis_prb96}
Tserkezis,~C.; Mortensen,~N.~A.; Wubs,~M. How nonlocal damping reduces plasmon-enhanced fluorescence in ultranarrow gaps. \emph{Phys. Rev. B} \textbf{2017}, \emph{96}, 085413\relax
\mciteBstWouldAddEndPuncttrue
\mciteSetBstMidEndSepPunct{\mcitedefaultmidpunct}
{\mcitedefaultendpunct}{\mcitedefaultseppunct}\relax
\EndOfBibitem
\bibitem[Yang \latin{et~al.}(2019)Yang, Zhu, Yan, Agarwal, Zheng, Joannopoulos, Lalanne, Christensen, Berggren, and Solja{\v{c}}i{\'{c}}]{Yang2019NatureFramework}
Yang,~Y.; Zhu,~D.; Yan,~W.; Agarwal,~A.; Zheng,~M.; Joannopoulos,~J.~D.; Lalanne,~P.; Christensen,~T.; Berggren,~K.~K.; Solja{\v{c}}i{\'{c}},~M. A general theoretical and experimental framework for nanoscale electromagnetism. \emph{Nature} \textbf{2019}, \emph{576}, 248--252\relax
\mciteBstWouldAddEndPuncttrue
\mciteSetBstMidEndSepPunct{\mcitedefaultmidpunct}
{\mcitedefaultendpunct}{\mcitedefaultseppunct}\relax
\EndOfBibitem
\bibitem[Yan \latin{et~al.}(2015)Yan, Wubs, and Mortensen]{Yan2015}
Yan,~W.; Wubs,~M.; Mortensen,~N.~A. Projected Dipole Model for Quantum Plasmonics. \emph{Phys. Rev. Lett.} \textbf{2015}, \emph{115}, 137403\relax
\mciteBstWouldAddEndPuncttrue
\mciteSetBstMidEndSepPunct{\mcitedefaultmidpunct}
{\mcitedefaultendpunct}{\mcitedefaultseppunct}\relax
\EndOfBibitem
\bibitem[Christensen \latin{et~al.}(2017)Christensen, Yan, Jauho, Solja\ifmmode \check{c}\else \v{c}\fi{}i\ifmmode~\acute{c}\else \'{c}\fi{}, and Mortensen]{Christensen_prl2017}
Christensen,~T.; Yan,~W.; Jauho,~A.-P.; Solja\ifmmode \check{c}\else \v{c}\fi{}i\ifmmode~\acute{c}\else \'{c}\fi{},~M.; Mortensen,~N.~A. Quantum Corrections in Nanoplasmonics: Shape, Scale, and Material. \emph{Phys. Rev. Lett.} \textbf{2017}, \emph{118}, 157402\relax
\mciteBstWouldAddEndPuncttrue
\mciteSetBstMidEndSepPunct{\mcitedefaultmidpunct}
{\mcitedefaultendpunct}{\mcitedefaultseppunct}\relax
\EndOfBibitem
\bibitem[Raza \latin{et~al.}(2011)Raza, Toscano, Jauho, Wubs, and Mortensen]{Raza2011}
Raza,~S.; Toscano,~G.; Jauho,~A.-P.; Wubs,~M.; Mortensen,~N.~A. Unusual resonances in nanoplasmonic structures due to nonlocal response. \emph{Phys. Rev. B} \textbf{2011}, \emph{84}, 121412(R)\relax
\mciteBstWouldAddEndPuncttrue
\mciteSetBstMidEndSepPunct{\mcitedefaultmidpunct}
{\mcitedefaultendpunct}{\mcitedefaultseppunct}\relax
\EndOfBibitem
\bibitem[Toscano \latin{et~al.}(2015)Toscano, Straubel, Kwiatkowski, Rockstuhl, Evers, Xu, Mortensen, and Wubs]{Toscano:2015}
Toscano,~G.; Straubel,~J.; Kwiatkowski,~A.; Rockstuhl,~C.; Evers,~F.; Xu,~H.; Mortensen,~N.~A.; Wubs,~M. Resonance shifts and spill-out effects in self-consistent hydrodynamic nanoplasmonics. \emph{Nat. Commun.} \textbf{2015}, \emph{6}, 7132\relax
\mciteBstWouldAddEndPuncttrue
\mciteSetBstMidEndSepPunct{\mcitedefaultmidpunct}
{\mcitedefaultendpunct}{\mcitedefaultseppunct}\relax
\EndOfBibitem
\bibitem[Yan(2015)]{Yan_prb2015}
Yan,~W. Hydrodynamic theory for quantum plasmonics: Linear-response dynamics of the inhomogeneous electron gas. \emph{Phys. Rev. B} \textbf{2015}, \emph{91}, 115416\relax
\mciteBstWouldAddEndPuncttrue
\mciteSetBstMidEndSepPunct{\mcitedefaultmidpunct}
{\mcitedefaultendpunct}{\mcitedefaultseppunct}\relax
\EndOfBibitem
\bibitem[Cirac\`{\i} and Della~Sala(2016)Cirac\`{\i}, and Della~Sala]{Ciraci:2016}
Cirac\`{\i},~C.; Della~Sala,~F. Quantum hydrodynamic theory for plasmonics: Impact of the electron density tail. \emph{Phys. Rev. B} \textbf{2016}, \emph{93}, 205405\relax
\mciteBstWouldAddEndPuncttrue
\mciteSetBstMidEndSepPunct{\mcitedefaultmidpunct}
{\mcitedefaultendpunct}{\mcitedefaultseppunct}\relax
\EndOfBibitem
\bibitem[Cirac\`{\i}(2017)]{Ciraci:2017}
Cirac\`{\i},~C. Current-dependent potential for nonlocal absorption in quantum hydrodynamic theory. \emph{Phys. Rev. B} \textbf{2017}, \emph{95}, 245434\relax
\mciteBstWouldAddEndPuncttrue
\mciteSetBstMidEndSepPunct{\mcitedefaultmidpunct}
{\mcitedefaultendpunct}{\mcitedefaultseppunct}\relax
\EndOfBibitem
\bibitem[Baghramyan \latin{et~al.}(2021)Baghramyan, Della~Sala, and Cirac\`{\i}]{Baghramyan:2021}
Baghramyan,~H.~M.; Della~Sala,~F.; Cirac\`{\i},~C. Laplacian-level quantum hydrodynamic theory for plasmonics. \emph{Phys. Rev. X} \textbf{2021}, \emph{11}, 011049\relax
\mciteBstWouldAddEndPuncttrue
\mciteSetBstMidEndSepPunct{\mcitedefaultmidpunct}
{\mcitedefaultendpunct}{\mcitedefaultseppunct}\relax
\EndOfBibitem
\bibitem[Lalanne \latin{et~al.}(2018)Lalanne, Yan, Vynck, Sauvan, and Hugonin]{Lalanne2018LigntInteraction}
Lalanne,~P.; Yan,~W.; Vynck,~K.; Sauvan,~C.; Hugonin,~J.-P. Light interaction with photonic and plasmonic resonances. \emph{Laser Photon. Rev.} \textbf{2018}, \emph{12}, 1700113\relax
\mciteBstWouldAddEndPuncttrue
\mciteSetBstMidEndSepPunct{\mcitedefaultmidpunct}
{\mcitedefaultendpunct}{\mcitedefaultseppunct}\relax
\EndOfBibitem
\bibitem[Yan \latin{et~al.}(2018)Yan, Faggiani, and Lalanne]{YanWei2018PRB}
Yan,~W.; Faggiani,~R.; Lalanne,~P. Rigorous modal analysis of plasmonic nanoresonators. \emph{Phys. Rev. B} \textbf{2018}, \emph{97}, 205422\relax
\mciteBstWouldAddEndPuncttrue
\mciteSetBstMidEndSepPunct{\mcitedefaultmidpunct}
{\mcitedefaultendpunct}{\mcitedefaultseppunct}\relax
\EndOfBibitem
\bibitem[Zhou \latin{et~al.}(2022)Zhou, Zhang, and Chen]{Zhou2022QSR}
Zhou,~Q.; Zhang,~P.; Chen,~X.-W. Quasinormal mode theory for nanoscale electromagnetism informed by quantum surface response. \emph{Phys. Rev. B} \textbf{2022}, \emph{105}, 125419\relax
\mciteBstWouldAddEndPuncttrue
\mciteSetBstMidEndSepPunct{\mcitedefaultmidpunct}
{\mcitedefaultendpunct}{\mcitedefaultseppunct}\relax
\EndOfBibitem
\bibitem[Zhou \latin{et~al.}(2021)Zhou, Zhang, and Chen]{Zhou:2021}
Zhou,~Q.; Zhang,~P.; Chen,~X.-W. General framework of canonical quasinormal mode analysis for extreme nano-optics. \emph{Phys. Rev. Lett.} \textbf{2021}, \emph{127}, 267401\relax
\mciteBstWouldAddEndPuncttrue
\mciteSetBstMidEndSepPunct{\mcitedefaultmidpunct}
{\mcitedefaultendpunct}{\mcitedefaultseppunct}\relax
\EndOfBibitem
\bibitem[Kamandar~Dezfouli \latin{et~al.}(2017)Kamandar~Dezfouli, Tserkezis, Mortensen, and Hughes]{Stephen2017NonlocalQNM}
Kamandar~Dezfouli,~M.; Tserkezis,~C.; Mortensen,~N.~A.; Hughes,~S. Nonlocal quasinormal modes for arbitrarily shaped three-dimensional plasmonic resonators. \emph{Optica} \textbf{2017}, \emph{4}, 1503--1509\relax
\mciteBstWouldAddEndPuncttrue
\mciteSetBstMidEndSepPunct{\mcitedefaultmidpunct}
{\mcitedefaultendpunct}{\mcitedefaultseppunct}\relax
\EndOfBibitem
\bibitem[Binkowski \latin{et~al.}(2019)Binkowski, Zschiedrich, Hammerschmidt, and Burger]{Binkowski2019ModalAnalysis}
Binkowski,~F.; Zschiedrich,~L.; Hammerschmidt,~M.; Burger,~S. Modal analysis for nanoplasmonics with nonlocal material properties. \emph{Phys. Rev. B} \textbf{2019}, \emph{100}, 155406\relax
\mciteBstWouldAddEndPuncttrue
\mciteSetBstMidEndSepPunct{\mcitedefaultmidpunct}
{\mcitedefaultendpunct}{\mcitedefaultseppunct}\relax
\EndOfBibitem
\bibitem[Raza \latin{et~al.}(2015)Raza, Bozhevolnyi, Wubs, and Mortensen]{raza_JPCM_2015}
Raza,~S.; Bozhevolnyi,~S.~I.; Wubs,~M.; Mortensen,~N.~A. Nonlocal optical response in metallic nanostructures. \emph{J. Phys.: Cond. Matter} \textbf{2015}, \emph{27}, 183204\relax
\mciteBstWouldAddEndPuncttrue
\mciteSetBstMidEndSepPunct{\mcitedefaultmidpunct}
{\mcitedefaultendpunct}{\mcitedefaultseppunct}\relax
\EndOfBibitem
\bibitem[Halevi(1995)]{halevi_prb51}
Halevi,~P. Hydrodynamic model for the degenerate free-electron gas: generalization to arbitrary frequencies. \emph{Phys. Rev. B} \textbf{1995}, \emph{51}, 7497--7499\relax
\mciteBstWouldAddEndPuncttrue
\mciteSetBstMidEndSepPunct{\mcitedefaultmidpunct}
{\mcitedefaultendpunct}{\mcitedefaultseppunct}\relax
\EndOfBibitem
\bibitem[Wegner \latin{et~al.}(2023)Wegner, Huynh, Mortensen, Intravaia, and Busch]{Wegner:2023}
Wegner,~G.; Huynh,~D.-N.; Mortensen,~N.~A.; Intravaia,~F.; Busch,~K. Halevi's extension of the Euler-Drude model for plasmonic systems. \emph{Phys. Rev. B} \textbf{2023}, \emph{107}, 115425\relax
\mciteBstWouldAddEndPuncttrue
\mciteSetBstMidEndSepPunct{\mcitedefaultmidpunct}
{\mcitedefaultendpunct}{\mcitedefaultseppunct}\relax
\EndOfBibitem
\bibitem[Yoon \latin{et~al.}(2019)Yoon, Selbach, Schumacher, Jose, and Schl\"{u}cker]{yoon_acsphot6}
Yoon,~J.~H.; Selbach,~F.; Schumacher,~L.; Jose,~J.; Schl\"{u}cker,~S. Surface plasmon coupling in dimers of gold nanoparticles: experiment and theory for ideal (spherical) and nonideal (faceted) building blocks. \emph{ACS Photonics} \textbf{2019}, \emph{6}, 642--648\relax
\mciteBstWouldAddEndPuncttrue
\mciteSetBstMidEndSepPunct{\mcitedefaultmidpunct}
{\mcitedefaultendpunct}{\mcitedefaultseppunct}\relax
\EndOfBibitem
\bibitem[Tserkezis \latin{et~al.}(2014)Tserkezis, Taylor, Beitner, Esteban, Baumberg, and Aizpurua]{tserkezis_part31}
Tserkezis,~C.; Taylor,~R.~W.; Beitner,~J.; Esteban,~R.; Baumberg,~J.~J.; Aizpurua,~J. Optical response of metallic nanoparticle heteroaggregates with subnanometric gaps. \emph{Part. Part. Syst. Charact.} \textbf{2014}, \emph{31}, 152--160\relax
\mciteBstWouldAddEndPuncttrue
\mciteSetBstMidEndSepPunct{\mcitedefaultmidpunct}
{\mcitedefaultendpunct}{\mcitedefaultseppunct}\relax
\EndOfBibitem
\bibitem[Tserkezis \latin{et~al.}(2014)Tserkezis, Herrmann, Valev, Baumberg, and Aizpurua]{tserkezis_oex22}
Tserkezis,~C.; Herrmann,~L.~O.; Valev,~V.~K.; Baumberg,~J.~J.; Aizpurua,~J. Optical response of threaded chain plasmons: from capacitive chains to continuous nanorods. \emph{Opt. Express} \textbf{2014}, \emph{22}, 23851--23860\relax
\mciteBstWouldAddEndPuncttrue
\mciteSetBstMidEndSepPunct{\mcitedefaultmidpunct}
{\mcitedefaultendpunct}{\mcitedefaultseppunct}\relax
\EndOfBibitem
\bibitem[Zhang \latin{et~al.}(2018)Zhang, Cai, Yu, Carregal-Romero, Parak, Sachan, Cai, Wang, Zhu, and Lei]{Lei2018}
Zhang,~Q.; Cai,~X.; Yu,~X.; Carregal-Romero,~S.; Parak,~W.~J.; Sachan,~R.; Cai,~Y.; Wang,~N.; Zhu,~Y.; Lei,~D.~Y. Electron Energy-Loss Spectroscopy of Spatial Nonlocality and Quantum Tunneling Effects in the Bright and Dark Plasmon Modes of Gold Nanosphere Dimers. \emph{Adv. Quantum Technol.} \textbf{2018}, \emph{1}, 1800016\relax
\mciteBstWouldAddEndPuncttrue
\mciteSetBstMidEndSepPunct{\mcitedefaultmidpunct}
{\mcitedefaultendpunct}{\mcitedefaultseppunct}\relax
\EndOfBibitem
\bibitem[P\'erez-Gonz\'alez \latin{et~al.}(2010)P\'erez-Gonz\'alez, Zabala, Borisov, Halas, Nordlander, and Aizpurua]{Aizpurua2010}
P\'erez-Gonz\'alez,~O.; Zabala,~N.; Borisov,~A.~G.; Halas,~N.~J.; Nordlander,~P.; Aizpurua,~J. Optical spectroscopy of conductive junctions in plasmonic cavities. \emph{Nano Lett.} \textbf{2010}, \emph{10}, 3090--3095\relax
\mciteBstWouldAddEndPuncttrue
\mciteSetBstMidEndSepPunct{\mcitedefaultmidpunct}
{\mcitedefaultendpunct}{\mcitedefaultseppunct}\relax
\EndOfBibitem
\end{mcitethebibliography}

\end{document}